\documentclass[prd,twocolumn,nofootinbib,aps,superscriptaddress,tightenlines,preprintnumbers]{revtex4}

\usepackage{slashed}
\usepackage{epsfig,latexsym,cancel,amssymb,amsmath,verbatim,mathrsfs}
\usepackage{color}
\usepackage{graphicx}
\usepackage{hyperref}

\usepackage{amssymb}
\usepackage{caption}
\usepackage{subcaption}

\definecolor{My_red}{cmyk}{0.00,1.00,1.00,0.20}

\usepackage{float}%

\def\L{\left(}
\def\R{\right)}
\def\wt{\widetilde}

\graphicspath{{images/}}

\usepackage{tikz-feynhand}

\begin{document}

\title{Confinement Bubble Wall Velocity via Quasiparticle Determination}

\author{Zhaofeng Kang}
\email[E-mail: ]{zhaofengkang@gmail.com}
\affiliation{School of Physics, Huazhong University of Science and Technology, Wuhan 430074, China}

\author{Jiang Zhu}
\email[E-mail: ]{jackpotzhujiang@gmail.com}
\affiliation{Tsung-Dao Lee Institute and School of Physics and Astronomy, Shanghai Jiao Tong University, 800 Dongchuan Road, Shanghai, 200240 China}
\affiliation{Shanghai Key Laboratory for Particle Physics and Cosmology, Key Laboratory for Particle Astrophysics and Cosmology (MOE), Shanghai Jiao Tong University, Shanghai 200240, China}
\affiliation{School of Physics, Huazhong University of Science and Technology, Wuhan 430074, China}

\date{\today}

\begin{abstract}
     
Lattice simulations reveal that the deconfinement-confinement phase transition (PT) of the hot pure $SU(N>2)$ Yang-Mills system is first order. This system can be described by a pool of quasigluons moving in the Polyakov loop background, and in this picture, we establish an effective distribution function for quasigluons, which encodes interactions among quasigluons and in particular the confinement effect. With it, we made the first attempt to calculate the confinement bubble wall velocity $v_w$ at the microscopical level, and we obtained a small velocity $v_w\sim 0.04$ using two different approaches, which is qualitatively consistent with others results like holography. 


\end{abstract}

\pacs{12.60.Jv,  14.70.Pw,  95.35.+d}

\maketitle

\newpage

\newpage

\section{Introduction}

In 1937, Landau proposed an approach to study continuous PT characterized by the spontaneous breaking of symmetry $G$. The key is to select an appropriate order parameter $\phi$ and construct an $G$-invariant Landau free energy $L [\phi,T]$, encoding the order of PTs and as well the behavior of relevant thermodynamic properties near the critical point. If the transition for the matter system is first order, the non-equilibrium evolution for phase conversion may be simplified to be the picture of bubble nucleation and growth~\cite{Langer:1969bc,Csernai:1992tj}. This coarse-grained theory is a combination of classical nucleation theory, dealing with liquid-gas PT, and the Ginzburg-Landau free energy. The order parameter is chosen to be the collective variable, the deviation of the energy density from its equilibrium value. When Landau's idea is applied to describe cosmic PTs, a big difference arises: the order parameter field here is a given ``matter" field whose VEV labels the vacuum, such as inflation field, Higgs field, etc., not just a macroscopic quantity selected in the thermodynamic matter system such as density or magnetization. The PT here is not related to changes in the structure of matter, but rather to changes in the vacuum of the field system, known as vacuum PT. It can proceed due to purely quantum fluctuations, and then first order phase conversion is through the vacuum decay developed by Coleman~\cite{Coleman:1977py} and Callan~\cite{Callan:1977pt}. Later, it is generalized to finite temperature by Affleck~\cite{Affleck:1980ac} and Linde~\cite{Linde:1981zj}. In this context, the order parameter does not necessarily transform under some symmetry, although most studies begin with a symmetry.


The electroweak PT (EWPT) is a well-known example to illustrate the difference. The EWPT is described by the transition of the Higgs field $\phi$, which couples to a gas of thermal particles, such as quarks and leptons etc, and gives them masses. In this cosmological field system, the Landau free energy $L[\phi,T]$ is usually not guessed at the mean-field level but can be calculated perturbatively. It generally takes the form $V_{eff}(\phi)=V_0(\phi)+V_T(\phi)$ with $V_0$ the zero temperature potential and $V_T(\phi)$ due to the plasma particles interacting with $\phi$. The resulting PT makes the universe transit from the EW symmetric vacuum to the broken one, but the plasma particles simply become heavier in the new vacuum without structural changes.

The QCD PT is much different than EWPT due to the confinement at infrared. This non-perturbative effect renders it is much more complicated to study the QCD PT, and we have not established a confirmative picture on it yet. In this work, we will consider the simpler case, the deconfinment-confinment PT (D-C PT) in $SU(N)$ pure Yang-Mills theory with $N>2$, a presumptive first order PT from the quasigluon (Debye screening) phase to the glueball (confinement) phase~\cite{Polyakov:1978vu, Susskind:1979up, Olive:1980dy, Witten:1984rs}. Most of the current approaches on D-C PT are grounded on the Polyakov loop model, which attempts to obtain the Landau free energy of the order parameter, the traced Polyakov loop $\ell$ defined in Eq.~(\ref{TPL}), without paying attention to the plasma. Therefore, they do not provide a way to calculate the phase expansion velocity, namely the bubble wall velocity $v_w$. This is a crucial quantity in predicting the stochastic gravitational wave background sourced by the first order D-C PT, which may be the only probe to the dark colored sector highly decoupled with the standard model. 



In this work, we will calculate $v_w$ based on the massive Yang-Mills theory with a temperature dependent quasigluon mass $M_g(T)$, which is assumed to ``absorb" the strong interaction~\cite{Kang:2022jbg} thus enabling us to perturbatively calculate $L[\ell,T]$ as in EWPT. Moreover, it keeps the picture of particle, and therefore it stands a chance to calculate the bubble wall velocity from a microscopic level for the first time. Actually, the key of our determination of $v_w$ in the D-C PT is to introduce quasi particle and moreover to find its distribution function as a non-ideal gas. Then, we find that the bubble velocity for this D-C PT is fairly small $\mathcal{O}(v_w)\leq 0.1$, qualitatively consistent with the results calculated using the fluid or holography approach~\cite{Kajantie:1986hq,Ignatius:1993qn,LiLi:2023dlc,Wang:2023lam,Bea:2021zsu,Bigazzi:2021ucw,Janik:2022wsx}. This means that the previous works taking $v_w=1$~\cite{Schwaller:2015tja,Helmboldt:2019pan,Huang:2020crf,Halverson:2020xpg,Kang:2021epo,Reichert:2021cvs,Morgante:2022zvc,Wang:2022gtn,Fujikura:2023fbi,Fujikura:2023lkn,Pasechnik:2023hwv} may overestimate the gravitational wave signature way much.




This paper is organized as follows. In Section~\ref{Sec:PTdyna}, we first briefly review the model to discuss the D-C PT for pure Yang-Mills theory and establish the corresponding statistical interpretations. Then, we discuss bubble dynamics and wall velocity of D-C PT in the early universe in the following section. Finally, conclusions and discussions, as well as three Appendices, are presented in the remaining two sections.

\section{{D-C PT: non-perturbative gluonic vacuum reconstruction}}\label{Sec:PTdyna}
Although there is no obvious order parameter for the D-C PT. The modulus of the expected value of the gauge invariant Polyakov loop (PL) in finite temperature($T$) theory may help us to understand the confinement
\begin{equation}\label{TPL}
    \ell(\vec {x})\equiv\frac{1}{N} {\rm tr}_c{\cal P}\exp{ig\int_0^\beta d\tau A_4(\vec {x},\tau)},
\end{equation}
where $\beta=1/T$ and $N$ is the color number, ${\rm tr}_c$ is the trace of the color index, ${\cal P}$ is the path order operator, and $A_4$ is the Euclidean temporal component of gauge field $A_0$ in the finite temperature field theory. It may be related to the free energy of a single static quark $Q$ via $\Phi=|\langle \ell(\vec {x})\rangle|\propto e^{-F_Q/T}$~\cite{McLerran:1981pb}.
It can be interpreted as the probability of observing a single quark in the gluonic plasma, which will be supported in a different way later. Then $\Phi\to 0$ implies an infinite $F_Q$, signaling color confinement. In contrast, $\Phi\neq 0$ means a finite $F_Q$ thus deconfinement. Moreover, PL is charged under the global center $Z_N\subset SU(N)$, and thus $\Phi$ is widely used as the 
effective order parameter for D-C PT in pure Yang-Mills theory (PYM)
. However, the presence of dynamical fermions with nonzero $N-$ality violates $Z_N$, rendering $\Phi$ merely a pseudo-order parameter. PL is a nonlocal operator of the fundamental field $A_0$, which cannot be gauged away at finite $T$. Then, $A_0$ are nothing but real scalars in the adjoint representation of $SU(N)$, and its transition from the deconfinement vacuum with a vanishingly small condensate to the confinement vacuum with a large condensate drives the transition of PL~\footnote{This is a losing statement, because unlike $A_ 0=0 $ simply leads to deconfinement phase, the confinement phase imposes a strong condition on the eigenvalues of $A_0$, if $\Phi=0$ is due to vanishing $\ell$ rather than the average over different $Z_N$ domains.}. In this sense, the D-C PT is a vacuum PT associated with spontaneous breaking of $SU(N)$.

\subsection{Quasiparticle moving in the PL background} 

We need a model to describe the system at $T_c^+$ 
with $T_c$ the critical temperature
, where the non-perturbative effect becomes significant. The weakly interacting quasigluon picture with $T$-dependent mass assumed to ``absorb" the dominant gluon interaction can successfully account for thermodynamics down to $T_c$~\cite{Goloviznin:1992ws,Peshier:1995ty}. But this statistical model does not address PT. Alternatively, Pisarski treats QGP as a condensate of PL~\cite{Pisarski:2000eq}, without any particle excitations at all. A well-designed Landau free energy works well from a few $T_c$ down to $T_c$, including the D-C PT. Quasigluons moving on the PL background is a hybrid approach~\cite{Meisinger:2003id,Ruggieri:2012ny,Sasaki:2012bi,Alba:2014lda,Islam:2021qwh}. Following the spirit of quasi particles, it implies a possible perturbative field picture of hot PYM at $T_c^+$: one may introduce a $T$-dependent gluon mass parameter to the original PYM Lagrangian, ``absorbing" the main nonperturbative interactions, and then the Landau free energy can be perturbatively calculated~\cite{Kang:2022jbg}. 

The model in Ref.~\cite{Kang:2022jbg} is a generalization to the massive PYM model~\cite{Reinosa:2014ooa} to the case with a temperature dependent quasigluon mass $M_g(T)$~\footnote{Ref.~\cite{Kang:2022jbg} studied this thermodynamics of this model, and the temperature dependent mass $M_g(T)$ is fitted the lattice data by machine learning at good level, and find $M_g(T)$ is almost the same for all color number $N$.},
\begin{align}\label{def:la}
\mathcal{L}=-\frac{1}{2g^2}{\rm tr}(F_{\mu\nu}F^{\mu\nu})+\bar{D}_\mu \bar{c}^a D^\mu c^a\\\nonumber
+ih^a \bar{D}_\mu\hat{A}^{\mu,a}+\frac{1}{2}M^2_g(T)\hat A^a_{\mu} \hat A^{a,\mu},
\end{align}
with $c,\bar c$ and $h$ the Ghost fields, real Nakanishi-Lautrup field, respectively. 
One should notice that the Ghosts are still massless, since the lattice data do not show that the correlators of ghosts develop a massive pole. It is consistent with perturbative calculation in finite-temperature field theory~\cite{Laine:2016hma}.
$A_\mu$ is decomposed into a background $\bar{A}_\mu=\bar A_0^i H^i\delta_{0\mu}$ ($H_i$ the Cartan generators) plus massive fluctuations $\hat{A}_\mu$. In the Landau-DeWitt gauge $\bar D_\mu \hat A^\mu=0$ with $\bar{D}_\mu^{ab}=\partial_\mu\delta^{ab}+f^{acb}\bar A^c_\mu$, we integrate out the fluctuations to obtain the potential $\mathcal{V}_{eff}=\frac{1}{V}\log Z$ with
\begin{equation}\label{effective action}
\begin{split}
\log Z=VT\Bigg[3\int \frac{d^3p}{(2\pi)^3}\log \det\left( 1-\hat{L}_Ae^{-\frac{E_g}{T}} \right)
\\
-\int \frac{d^3p}{(2\pi)^3}\log \det\left( 1-\hat{L}_A e^{-\frac{|\vec{p}|}{T}} \right)\Bigg],
\end{split}
\end{equation}
where $E_g=\sqrt{|\vec{p}|^2+M_{g}^2}$ and $\hat{L}_A$ is the thermal Polyakov loop in the adjoint representation. The details of the derivation of this result can be found in the appendix of~\cite{Kang:2022jbg}. One may parameterize $\bar A_4=i\bar A_0=\frac{2\pi}{g\beta } {\rm diag}(q_1,...,q_N)$ with $\sum_{i-1}^{N}q_i=0$ for a real PL. Then $\hat{L}_A={\rm diag}[e^{\beta\mu_1},e^{\beta\mu_2},...,e^{\beta\mu_{N^2-1}}]$, where $\mu_a$ picks one value of $\{i2\pi q_{ij}/\beta\}$ with $q_{ij}\equiv q_i-q_j$. We take the assumption of equal eigenvalue distribution $q_{ij}=\frac{i-j}{N}s$~\cite{Roberge:1986mm,Dumitru:2010mj,Kang:2022jbg}. For example, in $SU(3)$ and $SU(4)$ they are respectively given by:
\begin{equation}\label{MEV}
\begin{split}
\frac{\beta}{2\pi}\mu_a=&\{0,i\frac{s}{3},-i\frac{s}{3},i\frac{2s}{3},-i\frac{2s}{3},i\frac{s}{3},-i\frac{s}{3},0\},
\\
\frac{\beta}{2\pi}\mu_a=&\{0,i\frac{s}{4},-i\frac{s}{4},i\frac{s}{2},-i\frac{s}{2},i\frac{s}{4},-i\frac{s}{4},0,
\\
& i\frac{s}{4},-i\frac{s}{4},i\frac{s}{2},-i\frac{s}{2},i\frac{3s}{4},-i\frac{3s}{4},0\}.
\end{split}
\end{equation}
They will be interpreted as an imaginary chemical potential. 
Take the expression of $A_4$ as a function of $s$ into Eq.~(\ref{TPL}), we will find that PL can also be expressed by $s$ as $l=\sin(\pi s)/[N\sin(\pi s/N)]$.

\begin{figure}[htbp]
\centering 
\includegraphics[width=0.395\textwidth]{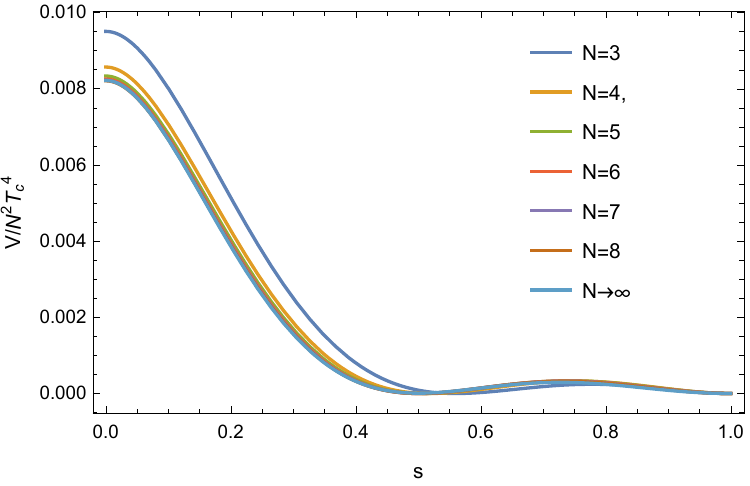} 
\caption{This figure comes from the Ref.~\cite{Kang:2022jbg}, it demonstrates the vacuum structure of the quasigluon model at the critical temperature $T_c$.} \label{Vacuum}
\end{figure}

Near $T_c$, the increasing quasigluon mass leads to the ghost contribution becoming relatively important. In this sense, it is a manifestation of non-perturbation effects in the infrared region, which destabilizes the perturbation deconfinement vacuum at $s_d\approx 0\ \&\ l\approx 1$~\footnote{One should notice that around the $T_c$, the deconfinement phase is located at $0<l<1$, and only reach $l=1$ when $T\ll T_c$.}, to arrive the confinement vacuum at $s_c=1\ \&\ l=0$ via first order PT. We have demonstrated the corresponding vacuum structure for different color numbers $N$ in $T_c$ in Fig.\ref{Vacuum}.


As we have previously claimed, the D-C PT may be treated as a vacuum PT. As for the transition of quasigluons to the confinement phase, it is somewhat similar to EWPT. As a Landau approach, the above model does not provide the microscopic mechanism for color confinement, but in the following, we will see that statistically it is attributed to the imaginary potential $\mu$. Then, we can claim that quasigluons in two different phases simply have different $\mu$, a change also encoded in the dispersion relation of quasigluons. 
 
\subsection{Effective statistical description of non-ideal quasi-gluon}
Different from the traditional quasiparticle model which is a statistical model, simply treating the QGP as a pool of ``ideal" quasiparticle gas with temperate varying mass~\footnote{Actually, it is shown that the statistical ensemble of quasigluon with $M_g(T)$ is not an ideal fluid~\cite{Sasaki:2008fg}.}, the pressure calculated in the field theory $P=-{\cal V}_{eff}$ includes the contribution of ghost and as well the background field PL. Note that in statistics, it corresponds to the pressure of the ensemble without bulk motion, denoted as $P$ exclusively. 

It is illustrative to rewrite $P$ in a more familiar form. To that end, let us first note that, in Eq.~(\ref{effective action}) $\hat{L}_A$ is a diagonal matrix. So, the determinant inside the logarithmic function is reduced to the sum over the $N^2-1$ eigenvalues. Next, the following trick allows us to rewrite the integration of the logarithmic function as the following
\begin{align} \nonumber
    \int^\infty_{-\infty}\frac{dp_z}{e^{\frac{E-\mu}{T}}-1}\frac{p_z^2}{E}
&=2\int^\infty_{\sqrt{p_\perp^2+m^2}}\frac{dE}{e^{\frac{E-\mu}{T}}-1}p_z
\\ \nonumber
&=2T\int^\infty_{\sqrt{p_\perp^2+m^2}}p_zd\log(1-e^{-\frac{E-\mu}{T}}),
\end{align}\label{trick}
where we have used $E=\sqrt{p_\perp^2+p_z^2+m^2}$ with $p_\perp^2=p_x^2+p_y^2$. Further by partial integration, we get the identity $\int^\infty_{-\infty}\frac{dp_z}{e^{\frac{E-\mu}{T}}-1}\frac{p_z^2}{E}
=-T\int^\infty_{-\infty}dp_z\log(1-e^{-\frac{E-\mu}{T}})$. The integral in Eq.~(\ref{effective action}) is symmetric with respect to $x$, $y$ and $z$, and therefore we arrive at $\int^\infty_{-\infty}\frac{d^3\vec{p}}{(2\pi)^3}\log(1-e^{-\frac{E-\mu}{T}})=-\frac{1}{3T}\int\frac{d^3\vec{p}}{(2\pi)^3}\frac{1}{e^{\beta(E-\mu)}-1}\frac{|\vec{p}|^2}{E}$. Finally, we get 
\begin{equation}\label{P-v=0}
\begin{aligned}
    P&=\sum_{a=1}^{N^2-1}\int\frac{d^3\vec{p}}{(2\pi)^3}\frac{1}{e^{\beta (E-\mu_a)}-1}\frac{|\vec{p}|^2}{E}
    \\
    &-\frac{1}{3}\sum_{a=1}^{N^2-1}\int\frac{d^3\vec{p}}{(2\pi)^3}\frac{1}{e^{\beta (|\vec{p}|-\mu_a)}-1}\frac{|\vec{p}|^2}{|\vec{p}|}.
\end{aligned}
\end{equation}    
This pressure resembles a mixture of ideal gas of massive gauge bosons with three degrees of freedom, plus one massless ``ghost", thus carrying a minus sign from its unphysical role in the FP procedure. Note that conventional Bose-Einstein distribution functions are twisted by the imaginary chemical potential $\mu_a$, and thus the $f_{a}(E,\mu_a)$ extracted from Eq.~(\ref{P-v=0}) is complex. 
However, $\mu_a$ always appear in pairs, i.e., $\mu_a=ic$ paired with $\mu_a'=-ic$, which can be seen in Eq.~(\ref{MEV}). This pairing makes the thermal quantities real after summing over $a$.

A similar result is known before, such as in Ref.~\cite{Hidaka:2008dr}, but it is obtained via a direct generalization of ordinary distribution functions based on the modified dispersion relation as discussed below 
\begin{equation}\label{dispersion}
\begin{split}
    (p_0-\mu_a)^2=|\vec{p}|^2+M_g^2,\ \ \ \ \ \ \ \ (p_0-\mu_a)^2=|\vec{p}|^2.
\end{split}
\end{equation}
They can be seen from the covariant derivative on the adjoint fields: $i[(\partial_0-i[\bar A_0,.])\hat A_\mu]_{ij}\to (p_0-i2\pi q^{ij}/\beta)[\hat A_\mu]_{ij}$, and a more rigid derivation from the quadratic term of Eq.~(\ref{def:la}) can be found in the Appendix~\ref{appA}. 

\begin{figure}[htbp]
\centering 
\includegraphics[width=0.4\textwidth]{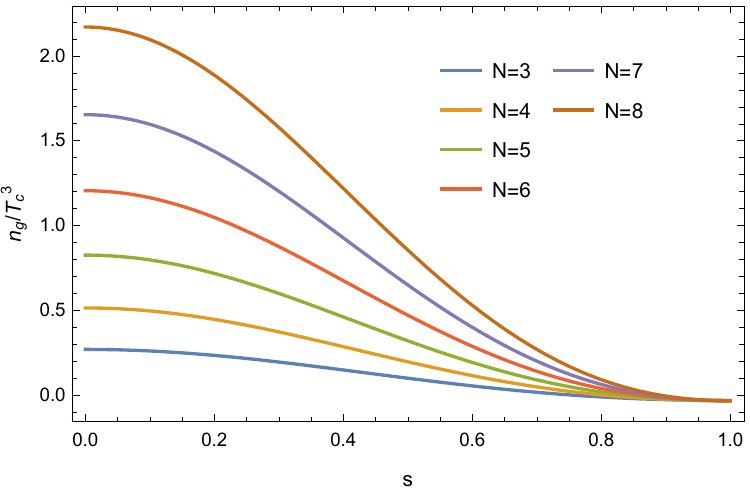} 
\caption{This plot demonstrates how the quasigluon number density evolved with $s$ at critical temperature $T_c$. One can find the number density drop when the order parameter $s$ is close to the confinement phase $s_c=1$.} \label{ng}
\end{figure}

Instead of changing the masses of plasma particles in the EWPT, here the background field enters the imaginary chemical potential of the quasigluons and results in statistical confinement. To see this, we calculate the number density of quasigluon using $f_{a}$ with $a$ summed over all color indices, 
\begin{equation}
n_g(s,T)=\sum_{i=1}^{N^2-1}\int\frac{d^3\vec{p}}{(2\pi)^3}\frac{1}{e^{\beta (E_i-\mu_i)}-1},
\end{equation}
in both phases. We show $n_g(T_c)$ as a function of $s$ in Fig.~\ref{ng}, to find that there is a significant suppression in the confinement phase.
It is a surprising result, because so far we have not yet forced quasigluons to form glueballs at $T_c^+$, the degree of freedom not included in our microscopic approach. A natural conjecture is that the quasigluons have been converted to glueballs during D-C PT. There is an interesting interpretation in the dynamical quasigluon scenario~\cite{Cassing:2007yg}: the force between quasigluons derived from the mean field is van der Waals like, and it becomes attractive as the quasigluon density drops below some critical value. Whether this mechanism can work here still needs further research, but we can also consider an alternative solution: the gluons may convert to vacuum condensate rather than to glueballs, which are produced via the wall oscillation after bubble collision. 
Anyway, for the purpose of calculating wall velocity, it is reasonable to speculate that it only plays a minor role, as lattice data tells us that the confinement phase is almost pressureless. 

\section{Bubble Dynamics and wall velocity of the D-C PT}\label{Sec:Wall}

Before proceeding with the calculation of the bubble wall velocity of the D-C PT, the nucleation temperature $T_n$ should be determined. The bubble nucleation rate of the confinement phase can be computed in the vacuum tunneling theory as $\Gamma=Ae^{-B}$, with $B$ given by
\begin{equation}\label{BT}
B(T)=4\pi\beta\int_0^{\infty} R^2dR \left[\frac{1}{2}\L\frac{d\phi_b}{dR}\R^2+\mathcal{V}_{eff}(\phi_b,T)\right],
\end{equation}
where $\mathcal{V}_{eff}$ is given in Eq.~(\ref{effective action}), and $\phi_b$ is the solution of the bounce equation for $\phi=T s$:
\begin{equation}
    \phi''+\frac{2}{r}\phi'=\frac{\partial \mathcal{V}_{eff}}{\partial\phi},\quad\phi'(0)=0,\quad \lim_{r\rightarrow\infty}\phi(r)=T_c s_{d}.
\end{equation}
 The solution can be obtained through the Mathematica package FindBounce~\cite{Guada:2020xnz} and then is plugged in Eq.~(\ref{BT}). The expression of $\mathcal{V}_{eff}$ is very complicated, so in parameter scanning, we take a sufficiently good analytical approximation to it~\cite{Kang:2022jbg}. Then, the nucleation temperature is determined by the nucleation condition in the radiation dominant universe as $B(T_n)\sim 140$. The model parameters and the corresponding nucleation temperatures can be found in Table~\ref{table-1}. With $T_n$, we will try to calculate the velocity of the bubble wall in the following.
\begin{table}[htbp] 
\centering
\scalebox{0.9}{
\begin{tabular}
{ |p{1.5cm}||p{1.2cm}|p{1.2cm}|p{1.2cm}|p{1.2cm}|p{1.2cm}|p{1.2cm}|   } 
 \hline
Color number& $N=3$ & $N=4$ &$N=5$&$N=6$&$N=7$&$N=8$\\
 \hline
$M_g(T_c)/T_c$& 2.7499 &2.7203& 2.7126&2.7099&2.7088&2.7083\\ 
 \hline
$dM_g/dT_c$& -5.7727 &-7.9951& -9.2891&-10.0965&-10.6261&-10.9954\\
 \hline
$T_n/T_c$& 0.999500 &0.999574& 0.999649&0.999699&0.999736&0.999763\\
 \hline
\end{tabular}
}
\caption{The quasigluon mass and corresponding nucleation temperature for the pure Yang-Mills Theory, the quasigluon mass we used when $T<T_c$ is getting from $M_g(T)=M_g(T_c)+(T-T_c)dM_g/dT_c$.}  
\label{table-1}
\end{table}

\subsection{Fluid pressure on the wall: from Micro to Statistical}

Due to the driving force of the two-phase vacuum pressure difference $\Delta P=\Delta V_{eff}(T=0)$, overcritical bubbles will maintain accelerated expansion in the fluid until the reaction force of the fluid increases to $F_{w}(v_w)/A= \Delta P$, resulting in a balanced force on the bubble wall and reaching a steady state. For EWPT, it is clear that the zero temperature scalar potential difference plays the role of driving force, and mass increasing when plasma particles cross the wall exerts an effective friction on the wall, impeding the bubble expansion. But in the D-C PT, we do not have a tree-level potential and then the steady equation is simply reduced to $F_w(v_w)/A=0$. 
We put a very detailed discussion in the Appendix.~\ref{CompEWDEC} to explain the above statement.

In principle, $F_w(v_w)$ can be calculated from a microscopic perspective. Consider a sufficiently large bubble expanding radially (aligned with the $z$-axis), and the bubble wall can be treated as a plane, namely the $xy$-plane. In the wall frame, suppose that a particle passing through a wall exerts a force $F_{sp}(z)$ on the wall, then the pressure acting on the bubble wall is the total force per unit area $\frac{F_w}{A}=\int dz \int \frac{d^3\vec{p}}{(2\pi)^3}F_{sp}(z)f(p,z)$~\cite{Moore:2000wx}. However, here $F_{sp}$ is hard to find, and thus we bypass this issue through the Boltzmann equation~\cite{Konstandin:2014zta,Hindmarsh:2020hop}
\begin{equation}\label{BE}
\begin{aligned}
    (p^\mu\partial_\mu+mF^\mu\partial_{p^\mu})\theta(p^0)\delta(p^2+m^2)f=C[f]
\end{aligned}
\end{equation}
with $F^\mu$ the external force acting on the particle. 

First, multiply $p^\nu$ on both sides of Eq.~(\ref{BE}) and integrate over $\vec p$, to find that the collision term vanishes as a result of energy momentum conservation in local thermal equilibrium. Further integrating over $p_0$, assuming that $F^\mu$ is momentum independent, we get 
\begin{equation}\label{T-conservation}
\begin{aligned}
        \partial_\mu\int\frac{d^3\vec{p}}{(2\pi)^3}\frac{p^\nu p^\mu}{E}f\equiv \partial_\mu T_f^{\mu\nu}=\int\frac{d^3\vec{p}}{(2\pi)^3}mF^\nu\frac{f}{E}.
\end{aligned}
\end{equation}
It gives the evolution of the energy tensor of the fluid system $\partial_\mu T_f^{\mu\nu}\neq 0$, due to the interaction between the fluid and the wall. To study steady bubble expansion, it is convenient to use Eq.~(\ref{T-conservation}) in the bubble wall frame where the time variable becomes irrelevant. Setting $\nu=z$ and using $F^z(z)=\frac{dp^z}{d\tau}=\frac{dt}{d\tau}\frac{dp^z}{dt}=\frac{E}{m}F_{sp}(z)$, Eq.~(\ref{T-conservation}) turns out to be $\frac{d}{dz}\int\frac{d^3\vec{p}}{(2\pi)^3}\frac{p_z^2}{E}f_w(p,z)=\int\frac{d^3\vec{p}}{(2\pi)^3}F_{sp}(z)f_w(p,z)$, with $f_w$ the distribution function in the bubble wall frame. Integrate it over $z$  we finally arrive
\begin{equation}\label{Tzz}
\Delta T^{zz}_f=\Delta \left[\int \frac{d^3\vec{p}}{(2\pi)^3}\frac{p_z^2}{E}f_w(p,z)\right]_{z=-\infty}^{z=\infty}= \frac{F_w}{A}.
\end{equation}
Through the steady-state distribution function on both sides, it establishes the relation between the pressure on the wall and the discontinuity of $T_f^{zz}$. That is, one can use this statistical formula to compute the pressure without knowing $F_{sp}$, as long as we can find the distribution functions in the two phases far from the wall. In reality, this is not a novel result and has been used before as a direct definition of pressure in the bubble frame~\cite{BarrosoMancha:2020fay}. Here we derived it from BE. In the following, we will use this formula to study the bubble wall velocity in the D-C PT.

Before proceeding, we would like to make a comment. An important assumption we should make is that the fluid inside (at least sufficiently far from the wall) the bubble has been thermalized, as is reasonable for slow-moving bubbles and fast scattering of the particle composition of the fluid. This is just the situation here. Otherwise, nonequilibrium effects will play a significant role that can not be addressed by the above approach and we may need to solve the BE just like in  EWPT~\cite{DeCurtis:2022hlx,Laurent:2022jrs,DeCurtis:2022llw}.

\subsection{Steady confinement bubble wall velocity}\label{sec:F}
The ordinary fluid approach~\cite{Steinhardt:1981ct,Espinosa:2010hh,Ai:2021kak,LiLi:2023dlc,Wang:2023lam,Ai:2024shx}, which expresses $T^{zz}_f$ in terms of the macroscopic thermodynamic quantities, is developed to deal with EWPT and does not apply here because the quasi-gluons significantly deviate from the ideal gas. But Eq.~(\ref{Tzz}) is based on the statistic approach, thus in principle applying to any fluid. Here, the crucial distribution function $f_w$ can be obtained by boosting the effective distribution functions (at infinity) in the fluid rest frame to the bubble frame
\begin{equation}\label{effective:dis}
f_{w,a}(v_w,p,z)=1/(\exp\beta[\gamma(E-v_w p_z+\mu_a(z))]-1),
\end{equation}
where the bubble expansion velocity $v_w$ enters.

\begin{figure}[htbp]
\centering 
\includegraphics[width=0.4\textwidth]{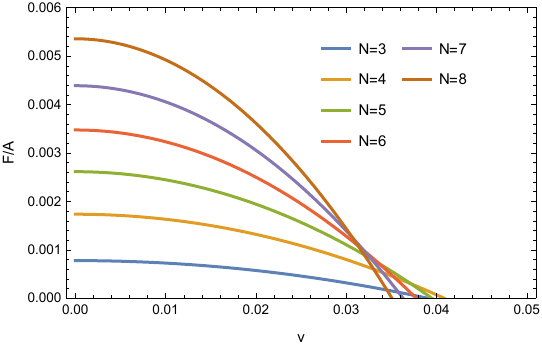} 4\caption{Steady bubble wall velocity for different color number from the statistic approach at the nucleation temperature $T_n$.} \label{F-v}
\end{figure}
In this frame, the steady bubble expansion means the static equation $F_w/A=\Delta T^{zz}_f=0$. 
Substituting Eq.(~\ref{effective:dis}) into the static equation Eq.~(\ref{Tzz}), we get the pressure acting on the wall, $F_w/A=P(s_c,v)-P(s_d,v)$ which is
\begin{equation}\label{finalpre}
\begin{aligned}
        \frac{F_{w}}{A}&=3\sum_{i=1}^{N^2-1}\int\frac{d^3\vec{p}\ \ p_z^2}{(2\pi)^3 E}[f^{pg}_{w,a}(v,p,\mu_{c,a})-f^{pg}_{w,a}(v,p,\mu_{d,a})]
        \\
        & +\sum_{i=1}^{N^2-1}\int\frac{d^3\vec{p}\ \ p_z^2}{(2\pi)^3 |\vec{p}|}[f^{g}_{w,a}(v,p,\mu_{d,a})-f^{g}_{w,a}(v,p,\mu_{c,a})]
\end{aligned}
\end{equation}
where $f^{pg/g}_{w,a}$ are distribution function for quasigluon and ghost in the wall frame, respectively.
Note that the distribution functions with an imaginary part seem to be unphysical, however, in various physical quantities, their integral in the momentum space leads to real values by virtue of the exact cancelation between the $\mu_i$ and $-\mu_i$ terms. We have calculated $F_w(v_w)/A$ for $N=3-8$ and plot them in Fig.~\ref{F-v}~\footnote{In this computation, we have set the temperature in such side of the bubble equal. This would not affect the general result if the bubble velocity is not very large.}. From this figure, it is clearly seen that, starting from a small initial velocity, the (decreasing) total force continues to accelerate the expansion of the wall until the wall velocity increases to a certain terminal value $v_w$. The terminal velocity for different $N$ takes a similar value, $v_w\approx 0.04$, which is qualitatively consistent with the previous results from the fluid or holography approach~\cite{Kajantie:1986hq,Ignatius:1993qn,LiLi:2023dlc,Wang:2023lam,Bea:2021zsu,Bigazzi:2021ucw,Janik:2022wsx}.

\subsection{Back to Micro}
There is a more widely used formula to calculate the microscopic friction exerted by plasma particles on the moving wall, not based on the specific external force, but on momentum transfer \cite{Arnold:1993wc,Bodeker:2009qy,Bodeker:2017cim,Azatov:2020ufh,Gouttenoire:2021kjv} :  
\begin{equation}
    \frac{dF_w}{A}=\frac{d^3\vec{p}}{(2\pi)^3}v_z f_w(p)\Delta p_z P(\Delta p_z)
\end{equation}
where $\frac{d^3\vec{p}}{(2\pi)^3}v_z f_w(p)$ is the flux of particle with velocity $v_z=p_z/\sqrt{|\vec{p}|^2+M_g^2}$,  and $P(\Delta p_z)$ is the probability for a quasigluon and ghost passing through the wall and transferring momentum $\Delta p_z$ to it.


In our approach, momentum transfer is induced by the change of $\mu_a$ in two phases, and can be estimated via the modified dispersion relation on the background in Eq.~(\ref{dispersion}).
In the bubble wall frame, due to translation symmetries, the conservation of energy and the perpendicular momentum are still valid and therefore for a quasigluon passing the wall we have $p_0^c=p_0^d$ and $p_\perp^c=p_\perp^d$, where the index $d/c$ label the De/Confinement phase. Then, one obtains the transfer $\Delta p_z^a=\sqrt{(p_0-\mu^a_d)^2-|\vec{p}|^2-m^2}  -\sqrt{(p_0-\mu^a_c)^2-|\vec{p}|^2-m^2}$ with $m=M_g$ or $0$. Now, 
\begin{align}
  \frac{F_w}{A}&=\int\frac{d^3\vec{p}_d}{(2\pi)^2}v^d_z f_w(p^d)[2p_z^d\mathcal{R}+(p_z^d-p_z^c)\mathcal{T}]\Theta(p_z^d)
  \\
  \nonumber
  &-\int\frac{d^3\vec{p}_c}{(2\pi)^2}v^c_z f_w(p^c)[2p_z^c\mathcal{R}+(p_z^c-p_z^d)\mathcal{T}]\Theta(-p_z^c),
\end{align}
where the distribution function is still given by Eq.~(\ref{effective:dis}) and 
$\mathcal{R}$/$\mathcal{T}$ is the reflection/transmission coefficient which an be computed in the thin wall approximation
. Again, we consider a nonrelativistic thin bubble wall and then expand $f_w/A$ in terms of $v_w$, to the linear order,
\begin{equation}\label{TFs}
\begin{split}
    \frac{F_w}{A}&=P(s_d)-P(s_c)+
    \\
    &2v_w\beta\int\frac{d^2p_\perp}{(2\pi)^2}\int^{\Re[{p_z'}]}_{0}\frac{dp_z^d}{2\pi}f(p_0)[f(p_0)+1]{p_z^d}^2+
    \\
    \nonumber
    &v_w\beta\int\frac{d^3\vec{p}^d}{(2\pi)^3}\frac{p_z^d}{\sqrt{|\vec{p}^d|^2+M_g^2}}f(p_0)[f(p_0)+1](p_z^d-p_z^c)^2,
\end{split}
\end{equation}
where, $p_z'=\sqrt{(m+\mu_c-\mu_d)^2-\vec{p}_\perp^2-m^2}$ and $\Re$ represents the real part. The second term comes from the reflection of particles, and we tend to ignore its contribution, the reason is given in the appendix. Because of this, in Fig.(\ref{F-v}), we only plot the contributions ignoring this term, which results in 
the changing of $v_w$; see Fig.~\ref{F-vm} in the Appendix. However, although considering the reflection effect will change the result of $v_w$, but the value of $v_w$ is still very small.
Note that in order to correctly recover $\Delta V_T$ in $v_w=0$ (the first term), we have to similarly take into account the contribution from the ghost ``particles". The numerical result is shown in the Fig.(\ref{F-vm}).
\begin{figure}[htbp]
\centering 
\includegraphics[width=0.4\textwidth]{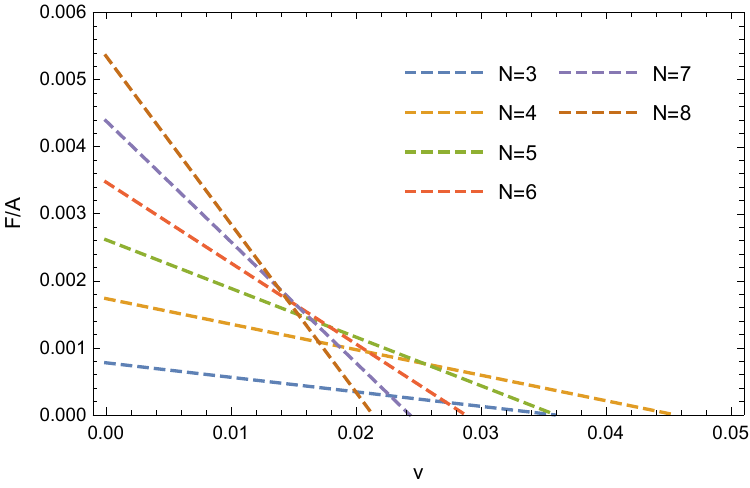} 
\caption{Steady bubble wall velocity for different color number from the microscopic approach without the reflection contributions at the nucleation temperature $T_n$.} \label{F-vm}
\end{figure}
As a remarkable consistency with the previous approach, we here also get $v_w$ around $0.04$ for different color numbers, despite the slight different shapes of $P(v)$ and $f_w(v)/A$ caused by the leading order approximations.

\section{Conclusions and Discussions}\label{Sec:Con}
Based on the picture of quasi-gluons moving in the PL background, we for the first time built a unified framework to study the dynamics of D-C PT, especially the bubble wall velocity. Using two different approaches, we both find $v_w\sim{\cal O}(0.01)$.

A very low bubble wall velocity $v_w\ll 1$ means that previous works taking $v_w=1$~\cite{Schwaller:2015tja,Helmboldt:2019pan,Huang:2020crf,Halverson:2020xpg,Kang:2021epo,Reichert:2021cvs,Morgante:2022zvc,Wang:2022gtn,Fujikura:2023fbi,Fujikura:2023lkn,Pasechnik:2023hwv} may overestimate the GW signature a lot. Moreover, for the D-C PT, typically $\wt \beta$ is huge, that is, the bubble nucleation rate is very fast. However, the slow-moving bubbles take a long time to occupy the whole universe. Then a problem arises: can the bubble collide with another bubble that reaches the terminal velocity rather than encounter another accelerating or nucleating bubble? This is a question that needs to be answered by numerical simulation. 

Our discussion can be applied to the case including dynamical quarks, which also move on the PL background. But there the situation becomes more complicated due to the interplay with chiral symmetry breaking and then the bubble wall is made up of both PL and $\sigma$ backgrounds, and as a result the friction receives contributions both from imaginary chemical potential and mass change~\cite{inworking}.

\noindent {\bf{Acknowledgements}}

This work is supported in part by the  National Key Research and Development Program of China Grant No.2020YFC2201504 and National Science Foundation of China (11775086).
 
\appendix
\section{Equation of Motion and Dispersion Relation}\label{appA}
In this appendix, we drive the EoMs and the dispersion relations for both quasi-gluon and ghost. We consider the case $N=3$, and the generalization to other color numbers is straightforward. The complete quadratic Lagrangian for quasi-gluon is given by
\begin{equation}
    \mathcal{L}=\frac{1}{2}(\partial_\rho\hat{A}^a_\mu\partial^\rho\hat{A}^{a,\mu}-C_{ab}^{\mu\nu}\hat{A}^a_\mu\hat{A}^b_\nu+2f_{abc}gf^{\mu\nu}\hat{A}^a_\mu\bar{A}^c_\rho\partial^\rho\hat{A}^b_\nu),\nonumber
\end{equation}
where the tensor $C_{ab}^{\mu\nu}$ reads
\begin{equation}
\begin{split}
    C_{ab}^{\mu\nu}=-\frac{1}{2}M_g^2g^{\mu\nu}\delta_{ab}-f_{abc}\partial^\nu\bar{A}^{\mu}_c&+2f_{abc}\bar{F}^{c,\mu\nu}
    \\
    &+f_{abd}f_{cbe}g^{\mu\nu}\bar{A}^d_\rho\bar{A}^{e,\rho}.\nonumber
\end{split}
\end{equation}
The resulting EoMs for the quasi-gluon field is 
\begin{equation}
\begin{split}
        &(\partial^2-M_g^2)\hat{A}^a_\mu-2f_{abc}\bar{A}^c_\rho\partial^\rho\hat{A}^b_\mu+f_{acd}f_{cbe}\bar{A}_\rho^d\bar{A}^{e,\rho}\hat{A}^b_\mu
        \\
        &-f_{abc}(\partial_\rho\bar{A}^{c,\mu}g^{\rho\nu}+\partial_\rho\bar{A}^{c,\rho}g^{\mu\nu})\hat{A}^b_\nu++2f_{abc}\bar{F}^{c,\mu\nu}\hat{A}^b_\nu=0.
\end{split}
\end{equation}
For a constantly uniform background $\bar{A}^0$, any derivative acting on it vanishes. Then the above EOMs are reduced to
\begin{equation}
\begin{split}
    &(\partial^2-M_g^2)\hat{A}^a_\mu-2f_{abc}\bar{A}^c_0\partial_t\hat{A}^b_\mu+f_{acd}f_{cbe}\bar{A}_0^d\bar{A}^e_0\hat{A}^b_\mu=0.
\end{split}
\end{equation}
In the $SU(3)$ theory, these equations can be rewritten into a matrix equation in the color space: $XA=0$, where $A=(\hat{A}^1_\mu,\hat{A}^2_\mu,...,\hat{A}^8_\mu)$, and the non-vanishing element in operator matrix $X$ is given by
\begin{equation}
\begin{split}
    &X_{11}=X_{22}=-\partial^2+M_g^2+(\bar{A}^3_0)^2\ \ \ \ X_{12}=-X_{21}=2\bar{A}^3_0\partial_t
    \\
    &X_{44}=X_{55}=-\partial^2+M_g^2+(\frac{1}{2}\bar{A}^3_0+\frac{\sqrt{3}}{2}\bar{A}^8_0)^2
    \\
    &X_{45}=-X_{54}=-(\bar{A}^3_0+\sqrt{3}\bar{A}^8_0)\partial_t
    \\
    &X_{66}=X_{77}=-\partial^2+M_g^2+(\frac{1}{2}\bar{A}^3_0-\frac{\sqrt{3}}{2}\bar{A}^8_0)^2
    \\
    &X_{67}=-X_{76}=(\bar{A}^3_0-\sqrt{3}\bar{A}^8_0)\partial_t\ \ \ \ X_{33}=X_{88}=-\partial^2.
    \nonumber
\end{split}
\end{equation}
Unlike the ordinary  Higgs field background,  the background $\bar{A}$ not only induces the mass squared for the fluctuations coupling to it, but also leads to kinematic mixings between fluctuations along different color directions. 

One can diagonalize the matrix equation $XA=0$ by redefining the field as
\begin{equation}
\begin{split}
    &B^1_\mu=\frac{1}{\sqrt{2}}(\hat{A}^1_\mu+\hat{A}^2_\mu),\ \ \ \ B^2_\mu=\frac{1}{\sqrt{2}}(\hat{A}^1_\mu-\hat{A}^2_\mu)\ \ \ \ B^3_\mu=\hat{A}^3_\mu
    \\
    &B^4_\mu=\frac{1}{\sqrt{2}}(\hat{A}^4_\mu+\hat{A}^5_\mu),\ \ \ \ B^5_\mu=\frac{1}{\sqrt{2}}(\hat{A}^4_\mu-\hat{A}^5_\mu)
    \\
    &B^6_\mu=\frac{1}{\sqrt{2}}(\hat{A}^6_\mu+\hat{A}^7_\mu),\ \ \ \ B^7_\mu=\frac{1}{\sqrt{2}}(\hat{A}^6_\mu-\hat{A}^7_\mu),\ \ \ \ B^8_\mu=\hat{A}^8_\mu,
    \nonumber
\end{split}
\end{equation}
and then the matrix equation becomes $Y B=0$, where $Y$ now is a diagonal operator matrix with diagonal elements $Y_{aa}=(-i\partial_t+\mu_a)^2+\vec{\nabla}^2-M_g^2$, with $\mu$ for $SU(3)$ 
\begin{equation}
\begin{split}
    &\mu_1=-\mu_2=i\bar{A}^3_0,\ \ \ \ \mu_3=\mu_8=0,
    \\
    &\mu_4=-\mu_5=-\frac{i}{2}(\bar{A}^3_0+\sqrt{3}\bar{A}^8_0)
    \\
    &\mu_6=-\mu_7=\frac{i}{2}(\bar{A}^3_0-\sqrt{3}\bar{A}^8_0).
\end{split}
\end{equation}
So, instead of a mass term, the spin-0 background field $\bar A_0$ generates an imaginary chemical potential for the fluctuations coupling to it Lorentz covariantly; it is true when quarks are introduced. The EOMs for the $B_\mu^a$ fields are
\begin{equation}
    (-i\partial_t+\mu_a)^2B^a_\mu+\vec{\nabla}^2B^a_\mu-M_g^2B^a_\mu=0.
\end{equation}
Substituting the plane wave solution $B_\mu^a=e^{ik.x}\epsilon^a_\mu$ into this equation, one can find the dispersion relationship $(p_0-\mu_a)^2=|\vec{p}|^2+M_g^2$ used in the context.

For the ghost field, one can follow the same procedure to get a similar EOM $(-i\partial_t+\mu_a)^2c^a+\vec{\nabla}^2c^a-M_g^2c^a=0$, and it gives rise to the dispersion relation $(p_0-\mu_a)^2=|\vec{p}|^2$. For gauge fixing term in Eq.(\ref{def:la}) can be removed by redefinition of the gauge field leaving a minus massive contribution and positive massless contribution with 1-DoFs. For a more detailed discussion about the gauge fixing term one can see our previous work \cite{Kang:2022jbg}.

\section{Balance equation of the bubble wall}\label{CompEWDEC}

To demonstrate the force balance of the bubble, let us consider a system undergoing phase transition which is described by a scalar-perfect fluid coupled system (although the fluid around the D-C PT is not a perfect fluid; this system is enough to get the key idea). The total energy-momentum tensor for the scalar-fluid coupled system is given by
\begin{equation}\label{EMTPHI}
    T^{\mu\nu}=\partial^\mu\phi\partial^\nu\phi-g^{\mu\nu}\left[\frac{1}{2}(\partial\phi)^2-V(\phi)\right]+T_f^{\mu\nu},
\end{equation}
where $V(\phi)$ is the zero temperature tree-level potential, and $T_f^{\mu\nu}$ is the energy-momentum for fluid system. If the fluid is perfect fluid, then $T_f^{\mu\nu}$ can be written as
\begin{equation}
    T_f^{\mu\nu}=\omega_f u^\mu u^\nu-g^{\mu\nu} p_f.
\end{equation}
where $p_f$ and $\omega_f$ are the pressure and enthalpy of the fluid, respectively. Using the above expression, we can rewrite the total energy-momentum tensor as
\begin{equation}
    T^{\mu\nu}=\partial^\mu\phi\partial^\nu\phi+\omega_f u^\mu u^\nu-g^{\mu\nu}\left[\frac{1}{2}(\partial\phi)^2-V(\phi)+p_f\right].
\end{equation}
In the planer approximation, consider the stable bubble that has reached the terminal velocity. From the conservation of the total energy-momentum tensor $\partial_\mu T^{\mu\nu}=0$ for $\nu=z$, we get $\partial_z T^{z z}=0$. Then, integrating this equation by $\int dz$, we find
\begin{equation}
    \Delta(\omega_f\gamma^2v^2-p_f)=-\Delta V.
\end{equation}
Its right-handed side is the difference between the zero temperature effective potential $V$ in the two vacua, while the left-handed side is the difference in $T^{z z}_f$ of the fluid. On the other hand, the most general form of $T_f^{zz}$ can also be written by statistical definition as
\begin{equation}
    T_f^{\mu\nu}=\int\frac{d^3\vec{p}}{(2\pi)^3}\frac{p^\mu p^\nu}{E}f(p,v).
\end{equation}
After using Eq.~(\ref{Tzz}), we get
\begin{equation}\label{FBlens}
    -\Delta V=\Delta\int\frac{d^3\vec{p}}{(2\pi)^3}\frac{p_z^2}{E}f(p,v)=-\frac{F_w}{A}.
\end{equation}
It is nothing but the equation that indicates the force balance and the stable bubble. So, in bubble dynamics,  the zero-temperature effective potential provides the driving force while $\Delta T_f^{zz}$ gives the fluid friction.

\begin{figure}[htbp]
\centering 
\includegraphics[width=0.4\textwidth]{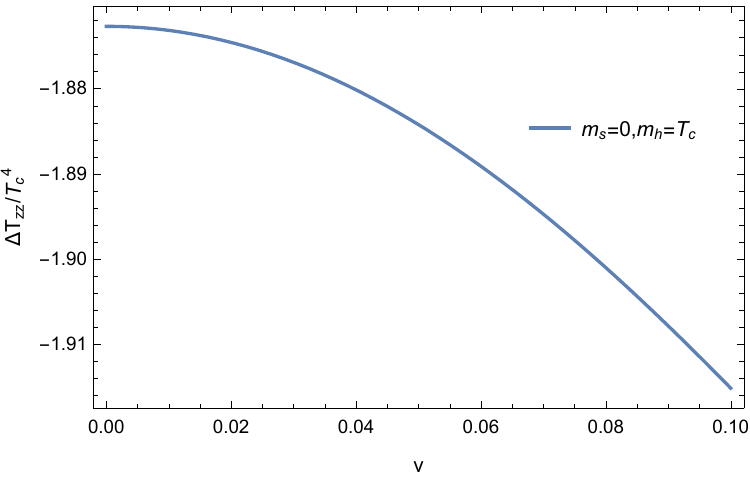} 
\caption{The electroweak friction force as a function of bubble velocity.}\label{FEW}
\end{figure}

Let us first consider a toy EWPT where the particles have mass $m_s=0$ and $m_h=T_c$ in the symmetric phase and the Higgs phase, respectively. Then we plot the right-hand side of Eq.~(\ref{FBlens}), that is, $\Delta T_{zz}$ normalized by $T_c^4$, as a function of $v$, and the result is shown in Fig.~\ref{FEW}. One can see that friction force increases monotonically with velocity and the reaching at terminal velocity happens at its balance with the nonzero $\Delta V$. 


However, in the quasigluon model~\cite{Kang:2022jbg}, there is no zero temperature effective potential for $\bar{A}_0$. One can see this by expanding the PYM Lagrangian $\mathcal{L}=-\frac{1}{4}F^a_{\mu\nu}F^{a,\mu\nu}$ by assuming $A_\mu=\bar{A}_\mu+\hat{A}_\mu$. Then, the tree-level zero-temperature effective Lagrangian for the classical background field $\bar{A}^a_\mu$ is determined by the Lagrangian for the zero-order fluctuation field $\mathcal{L}^0$. Assuming $\bar{A}^a_\mu=\bar{A}^a_0 g_{0\mu}$ with $\bar A_0^a$ coordinate independent, one gets
\begin{equation}
\begin{split}
    \mathcal{V}_{tree}=-\mathcal{L}^0&=\frac{1}{4}\bar{F}^a_{\mu\nu}\bar{F}^{a,\mu\nu}|_{\bar{A}^0=const}
    \\
    &=\frac{1}{4}f^{abc}f^{ade}\bar{A}_0^b\bar{A}_0^c\bar{A}_0^d\bar{A}_0^e
\end{split}
\end{equation}
So, we have $\mathcal{V}_{tree}=0$ since $b/d$, $c/d$ index in $f^{abc}/f^{ade}$ is anti-symmetric. In this case, the terminal velocity of the bubble is simply determined by $F_W/A=\Delta T^{zz}_f=0$.

\section{Detial Computation of Microscopic Friction Force}\label{appB}
In this appendix, we will show how to get Eq.(\ref{TFs}). Let us start with the general expression for the pressure
\begin{equation}
    \frac{dF}{A}=\frac{d^3\vec{p}}{(2\pi)^3}v_z f_w(p)\Delta p_z P(\Delta p_z)
\end{equation}
where $F$ representing the force acting one the wall, $\frac{d^3\vec{p}}{(2\pi)^3}v_z f_w(p)\Delta p_z$ representing the flux time the momentum transfer and $P(\Delta p_z)$ representing the probability for event that transfer momentum $\Delta p_z$. This is a widely used definition for the microscopic force, and also applicable at the level of quantum field theory, which includes any $1\to {\rm any}$ transition event. In this letter, we only consider the leading one at low wall velocity, the $1\rightarrow 1$ event. In this case, the event only contains reflection and transmission, and the above equation becomes
\begin{equation}
    \frac{dF}{A}=\frac{d^3\vec{p}^d}{(2\pi)^3}v_z^d f_w(p^d)[2p_z^d \mathcal{R}+(p_z^d-p_z^c)\mathcal{T}]
\end{equation}
where $\mathcal{R}$ and $\mathcal{T}$ are the reflection and transmission rate, respectively. In principle, they can be determined by solving the quantum mechanics problem. One can find that $\mathcal{R}=\frac{(p_z^d-p_z^c)^2}{(p_z^d+p_z^c)^2}$ and $\mathcal{T}=\frac{4p_z^dp_z^c}{(p_z^d+p_z^c)^2}$, for more detail one can see ref.\cite{Arnold:1993wc} or barrier tunneling problem from any textbook of quantum mechanics. 

The distribution function in the bubble wall frame is given by $f_w(p)=\frac{1}{\exp[\beta\gamma(p_0-v_w p_z)]-1}$, with the dispersion relation $p_0=\sqrt{|\vec{p}|^2+M_g^2}+\mu_a$. We expand this distribution function in terms of small $v_w$,
\begin{equation}
    f_w(p)\approx f(p_0)+v_w\beta p_z f(p_0)[f(p_0)+1],
\end{equation}
with $f(p_0)=1/[\exp(\beta\gamma p_0)-1]$. Then, the total force acting on the wall, considering the contributions from both sides,  can be written as
\begin{equation}\label{FPA}
\begin{split}
    \frac{F}{A}&=\frac{F_d}{A}-\frac{F_c}{A}
    \\
    &=\int\frac{d^3\vec{p}^d}{(2\pi)^3}v_z^d f_w(p^d)[2p_z^d\mathcal{R}+(p_z^d-p_z^c)\mathcal{T}]\theta(p_z^d)
    \\
    &-\int\frac{d^3\vec{p}^c}{(2\pi)^3}v_z^c f_w(p^c)[2p_z^c\mathcal{R}+(p_z^c-p_z^d)\mathcal{T}]\theta(-p_z^c)
    \\
    &\approx\int\frac{d^3\vec{p}^d}{(2\pi)^3}v_z^d [f(p_0)+v_w\beta p^d_z f(p_0)(f(p_0)+1)]
    \\
    &\times[2p_z^d\mathcal{R}+(p_z^d-p_z^c)\mathcal{T}]\theta(p_Z^d)-\int\frac{d^3\vec{p}^c}{(2\pi)^3}v_z^c[f(p_0)
    \\
    &+v_w\beta p_z^c f(p_0)(f(p_0)+1)][2p_z^c\mathcal{R}+(p_z^c-p_z^d)\mathcal{T}]\theta(-p_z^c).
\end{split}
\end{equation}
where we have used that $p_0$ is a conserved quantity $p_0^c=p_0^d=p_0$. At both sides of the bubble wall, the background field is a constant field and therefore, through the dispersion relation, we can rewrite $dp_z$ in terms of $dp_0$, 
\begin{equation}\label{p0pz}
    dp_0=\frac{p_z^vdp_z^v}{\sqrt{|\vec{p}^v|^2+M_g^2}},\quad v=c/d.
\end{equation}
Then, using  $v_z^v=p_z^v/\sqrt{|\vec{p}^v|^2+M_g^2}$, Eq.(\ref{FPA}) becomes
\begin{equation}
\begin{split}
    \frac{F}{A}&\approx\int\frac{d^2p_\perp}{(2\pi)^2}\int_{m+\mu_d}^\infty\frac{p_0}{2\pi}f(p_0)[2p_z^d\mathcal{R}+(p_z^d-p_z^c)\mathcal{T}]
    \\
    &-\int\frac{d^2p_\perp}{(2\pi)^2}\int_{m+\mu_c}^\infty\frac{p_0}{2\pi}f(p_0)[2p_z^c\mathcal{R}+(p_z^c-p_z^d)\mathcal{T}]
    \\
    &+v_w\beta\int\frac{d^2p_\perp}{(2\pi)^2}\int_{m+\mu_d}^\infty\frac{dp_0}{2\pi}f(p_0)(f(p_0)+1)p_z^d[2p_z^d\mathcal{R}
    \\
    &+(p_z^d-p_z^c)\mathcal{T}]+v_w\beta\int\frac{d^2p_\perp}{(2\pi)^2}\int_{m+\mu_c}^\infty\frac{dp_0}{2\pi}f(p_0)
    \\
    &\times(f(p_0)+1)p_z^c[2p_z^c\mathcal{R}+(p_z^c-p_z^d)\mathcal{T}],
\end{split}
\end{equation}
Assuming that particles with frequency in $[m+\mu_d,m+\mu_c]$ cannot enter bubbles, then the $v_w$ independent term gives
\begin{equation}
\begin{split}
    \frac{F_{ind}}{A}&=\int\frac{d^2p_\perp}{(2\pi)^2}\int_{m+\mu_d}^{m+\mu_c}\frac{dp_0}{2\pi}f(p_0)2p_z^d
    \\
    &+\int\frac{d^2p_\perp}{(2\pi)^2}\int_{m+\mu_c}^\infty\frac{dp_0}{2\pi}f(p_0)2[(p_z^d-p_z^c)(\mathcal{T}+\mathcal{R})]
    \\
    &=2\int\frac{d^2p_\perp}{(2\pi)^2}\left[\int_{m+\mu_d}^\infty \frac{dp_0}{2\pi}f(p_0)p_z^d-\int_{m+\mu_c}^\infty\frac{dp_0}{2\pi} f(p_0)p_z^c\right],
\end{split}
\end{equation}
with $\mathcal{R}+\mathcal{T}=1$. Now, using Eq.(\ref{p0pz}), one can prove that it indeed reproduces the difference of equilibrium pressure  Eq.(\ref{P-v=0}) as expected. However, we are still not able to understand the imposed condition to filter quasigluons.   

\begin{figure}[htbp]
\centering 
\includegraphics[width=0.4\textwidth]{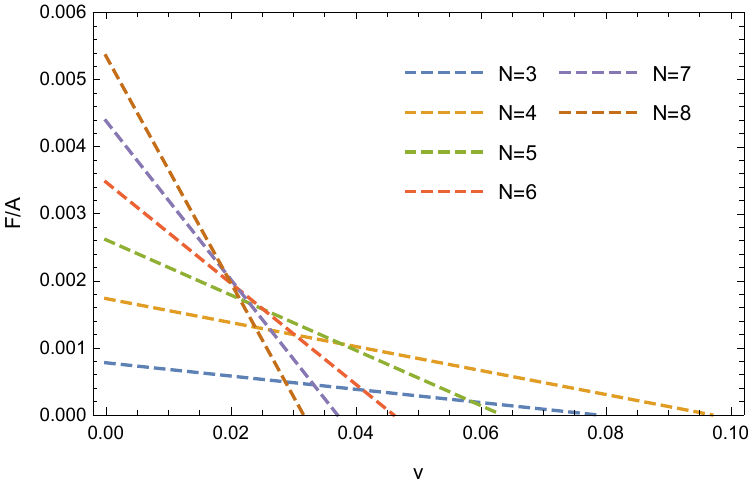} 
\caption{Steady bubble wall velocity for different color numbers from the microscopic approach with the reflection at the nucleation temperature $T_n$.}\label{F-vR}
\end{figure}

For the $v_w$ dependent part, we can do the same procedure and find that 
\begin{equation}
\begin{split}
    \frac{F_v}{A} &=2v_w\beta\int\frac{d^2p_\perp}{(2\pi)^2}\int_{m+\mu_d}^{m+\mu_c}\frac{dp_0}{2\pi}f(p_0)[f(p_0)+1]{p_z^d}^2
    \\
    &+2v_w\beta\int\frac{d^2p_\perp}{(2\pi)^2}\int_{m+\mu_c}^{\infty}\frac{dp_0}{2\pi}f(p_0)[f(p_0)+1](p_z^d-p_z^c)^2
    \\
    &=2v_w\beta\int\frac{d^2p_\perp}{(2\pi)^2}\int^{p_z'}_{0}\frac{dp_z^d}{2\pi}f(p_0)[f(p_0)+1]{p_z^d}^2
    \\
    &+v_w\beta\int\frac{d^3\vec{p}^d}{(2\pi)^3}\frac{p_z^d}{\sqrt{|\vec{p}^d|^2+M_g^2}}f(p_0)[f(p_0)+1](p_z^d-p_z^c)^2.
\end{split}
\end{equation}
where we have use the expression for $\mathcal{T}$ and $\mathcal{R}$ and $p_z'=\sqrt{(m+\mu_c-\mu_d)^2-\vec{p}_\perp^2-m^2}$. The first integral will diverge, owing to a complex upper limit of the integration. In order to get concrete numerical results, currently we are forced to artificially ignore the first term or just choose the real part of this upper limit $\Re[p_z']$. Eventually, we reach Eq.~(\ref{TFs}). To estimate the effect of this procedure, we show the bubble velocity in the cases both excluding and including the reflection contribution in Fig.~(\ref{F-vR}), from which one can see that the reflection term gives rise to a change in $v_w$.

\noindent

\ 
\ 
\ 

\

\end{document}